\documentclass[12pt]{article}

\usepackage[T2A]{fontenc}
\usepackage{graphicx}
\usepackage{amssymb,amsmath}
\usepackage{array}
\usepackage{booktabs}

\def\d{{\rm d}}
\newcommand{\p}{\partial}
\newcommand{\ve}{\varepsilon}
\def\({\left(}\def\){\right)}
\def\il{\int\limits}
\def\mlp{\max\limits_p}

\title{Galactic bar: \\normal mode of the stellar disk \\or superposition of transient spirals?}

\author{E.V. Polyachenko\\
Institute of Astronomy RAS, Moscow}
\date{}

\begin{document}
\maketitle

\section*{\LARGE Abstract}

{\footnotesize 
Several mechanisms of bar formation in stellar galactic disks are considered, including Toomre swing amplification and normal mode approach.
On example of the well-known model of Kuzmin--Toomre using N-body simulations it was shown that the stellar bar is developed as a result of the evolution of an
unstable normal mode. The pattern speed and the growth rate found agree well with linear perturbation theory. Nonlinear evolution of the bar is followed. Role of
the growing transient spirals in bar formation is discussed.}

\thispagestyle{empty}


\baselineskip = 20pt

\section{Introduction}

According to observations, bars are more common in interacting galaxies, but the percentage of occurrence of isolated galaxies with bar is also high.
Consequently, there must be an internal mechanism for the bar formation. Unlike spirals, which presumably require gas inflow from the outside
(Bournaud \& Combes 2002), long-lived bars are easily obtained in the simplest models of stellar disks (Hohl 1971). However, despite the extensive,
mainly numerical work, the generally accepted theory for bar formation does not still exist. To date, several mechanisms have been
proposed, two of which will be touched here: swing amplification theory of transient spirals (Toomre 1981), and the unstable precessing modes by Polyachenko (2004,
2005). See also Lynden-Bell (1996) for some other formation mechanisms.

One of the problems in the theory of spiral structures is the so called winding problem:  differentially rotating disk  
quickly winds up and dissolves material arms. The hypothesis that a pattern rotates without changing its shape due to gravity has been proposed for the first time
by Lindblad (1963, 1964), and then, in the context of stationary density waves, by Lin and Shu (1964, 1966). In the latter, a WKB-theory of tightly-wound spirals
has been proposed. Then Toomre (1969) shows that such spirals evolve: propagate inward and wind up like material arms.
So, instead of stationary density waves, the attention was switched to unstable density waves.

In the textbook by Binney and Tremaine (2008) which is now the standard in stellar dynamics courses, a physical interpretation of bar-mode instability based on
swing amplification mechanism is proposed. The mechanism is based on the effect of significant (tenfold) increase of the amplitude of spiral
waves in the corotation region during its transformation from leading to trailing. The trailing waves passing through the disk center can
become leading ones and re-enter the region of corotation. Thus multiple amplification and instability may occur. According to the theory, spiral patterns occur
when growth rates are large, while bars occur for much lower growth rates, when the amplitudes of the leading and trailing waves are nearly equal. Then the
interference pattern (in the linear regime when amplitudes are small) will have a lumpy structure with 90$^\circ$ spacing of the successive density maxima (see
Binney, Tremaine 2008, Fig.\,6.23).

There are alternative \textit{modal} theories in which observed structures are unstable normal modes of the galactic disk. In contrast to the WKB theory by
Lin\,-\,Shu, where the disk is considered locally, here a problem of finding of the global modes that span the entire disk or its substantial part is posed.

Note that calculation of the disk normal modes is a difficult task. Until recently, the only method was the Kalnajs matrix method (1977). 
Almost all available studies are adaptation of this method for a specific model. The method is characterized by a nonlinear equation for the
unknown oscillation frequency, which makes it difficult to find solutions.

Recently Polyachenko (2004, 2005) proposed a new method for the calculation of the normal modes, where the unknown oscillation frequencies are obtained from a
\textit{linear} eigenvalue problem. Solution of the problem are eigenvalues of a matrix, so there are no missing roots. In addition, this approach can be
generalized to more complex systems\footnote{A similar method for spheres was first given in Polyachenko et al. (2007); loss cone gravitational instability,
radial orbit instability, and generalized polytropic models were studied in Polyachenko et al. (2008\,--\,2011).} and allows the insight for the bar-formation
mechanism.

In Polyachenko (2004) the galactic bar is an unstable normal mode in the system of precessing orbits. The author deduces an equation for
normal modes away from resonances (except the inner Lindblad) and shows that stellar disks can support stable normal modes with pattern speeds above the
maximum of precession rate, localized in the central region of the disk. These normal modes become unstable when the ``outer'' resonances (the corotation, outer
Lindblad, etc.) are taken into account. 

Lynden-Bell and Kalnajs (1972) have shown that an outside disturbance, such as a satellite galaxy, cause the ILR stars to lose angular momentum, while all outer
resonances gain it\footnote{The same is true for the energy: the change of energy and angular momentum is related by $ \Delta E / \Delta L = 
\Omega_p $, where $\Omega_p $ is a pattern speed.}. When a growth rate of the perturbation is small, only stars on resonances interact with disturbance.
However, for larger growth rates, stars in adjacent regions of the phase space are involved in exchange of the angular momentum.

When the unstable mode develops in the isolated disk, a wave composed of stars in ILR region plays a role of the outside perturbation at outer resonances. At the
same time, ILR stars are influenced by a perturbation of the gravitational potential from stars on the outer resonances. The total angular momentum of the wave 
always vanishes by the conservation law.

\begin{figure}[t!]
\centerline{%
 \includegraphics[width=80mm, draft=false]{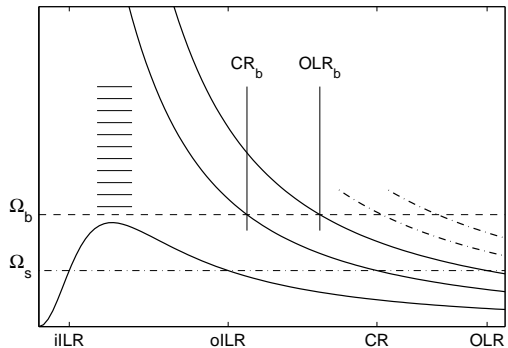}
 }
\caption{\footnotesize 
The frequency curves $ \Omega - \kappa / 2 $, $ \Omega $, $ \Omega + \kappa / 2 $ v.s. radius $R$ for a typical galaxy with slowly increasing
rotation curve (solid lines, bottom to top, respectively), where $ \Omega (R) $ and $ \kappa (R) $ --  angular and epicyclic frequencies of star oscillations.
$\Omega - \kappa / 2 $ determines the precession rate of stars on nearly circular orbits. The positions of the main resonances for two values of pattern speeds
$ \Omega_ {\textrm {s}} $ and $ \Omega_b $ of two-arm ($m = 2$) spiral pattern is shown. The intersection of $ \Omega_ {\textrm {s}} $  with curves
indicates the position of the inner Lindblad resonances (iILR, oILR), the corotation resonance (CR) and outer Lindblad resonance (OLR). At the higher speed 
$\Omega_b $ inner Lindblad resonances are absent. The intersections of the horizontal line $ \Omega_b $ with dashed-dotted curves show more distant resonances.
Horizontal lines above the precession maximum show the area of ``almost resonance'' for $ \Omega_b $, see text.}
 \label{fig_res}
\end{figure}

Fig.\,\ref{fig_res} shows the location of main resonances for a system with slowly increasing rotation curve (for a Plummer-like model, see Section 2) for 
two-arm ($m = 2$) spiral pattern and two pattern speeds $\Omega_{\textrm{s}}$  and $\Omega_{\textrm{b}}$. For $\Omega_{\textrm{s}}$ below maximum of the precession
curve  $\Omega - \kappa/2$, there are two inner Lindblad resonances (ILRs), the corotation resonance (CR), the outer Lindblad resonance (OLR), and further
resonances. For $\Omega_{\textrm{b}}$, the ILRs are absent.

The observed bars in galaxies have pattern speeds $\Omega_{\textrm{b}}$ slightly exceeding the precession curve maximum (Combes \& Elmegreen 1993, Lynden-Bell
1996).
A bar-mode with pattern speed $\Omega_{\textrm{b}}$ and small growth rate cannot exist, since the only source of the angular momentum for the outer resonances is
absent. However, for significant growth rates it is sufficient to have ``almost resonance'', from where the angular momentum can be effectively taken (in
Fig.\,\ref{fig_res} the area market by horizontal lines).

Thus, using numerical simulation, we can make a comparison of the two theories of the bar formation on several points:
\begin{description}
 \item[Process of bar formation.] If the bar is formed as a result of nonlinear evolution of an unstable normal mode, its formation can be easily tracked
using Fourier analysis of the density distribution. On the contrary, if the bar is formed from a lumpy structure that is a superposition of transient spirals, a
clear picture of the evolution of one mode should not be found.
\item[Growth rate in the linear regime.] In swing amplification theory, the growth rates of bar modes are low (several percents of the pattern speed). According to
the theory of unstable precessing modes, growth rates can not be small.
\item[Form of the pattern.] As a rule, a disk possesses several unstable modes. Evidently, the mode with the highest growth rate absorbs the others in the
non-linear regime. Therefore, the theory of unstable normal modes predicts that the bar is formed from a spiral with few density maxima and the highest
growth rate. On the contrary, swing amplification theory predicts lumpy structures made of superposition of the leading and trailing spirals.
 \end{description}

It makes sense to perform the comparison using a well-known model, such as Kuzmin-Toomre disk with Kalnajs (1976) distribution function (DF)  $m_\textrm{K}=6$,
$L_c=0.25$. This model was both studied numerically (Athanassoula, Sellwood 1986) and analytically (Polyachenko 2004, 2005; Jalali,
Hunter 2005). In the latter, frequencies of unstable modes and their patterns were found for several models, including the mentioned above.
It is worth comparing our numerical results with their linear perturbation analysis. Note that similar task for two-arm ($m=2$), and also for $m=3,4$
perturbations has been made by Khoperskov et al. (2007). The connection between the linear bar-mode and the well-developed bar can be established using power
spectra of unstable two-arm perturbations.

Section 2 describes the model, gives expressions for the construction of the equilibrium model in presence of softening, and discusses some of the features of
numerical simulations. In Section 3, for the  Kuzmin--Toomre model a comparison with the linear perturbation theory and analysis of power spectra for different
radii and spirality is presented. Section 4 resumes the results and discusses a role of the swing amplification mechanism in the bar
formation.

\section{Numerical simulation of a stellar disk}

\subsection{The dynamical model}

The simplest models for study of the large-scale bar formation are self-gravitating models with rotation curve slowly increasing with radius. A well-known example
is  Kuzmin--Toomre disk with the surface density profile
\begin{align}
\Sigma(R) = \frac{M_d}{2\pi a^2} \left(1 + \frac{R^2}{a^2} \right)^{-3/2}\quad,
\label{sd}
\end{align}
where $a$ is the radial scale, $M_d$ is a disk mass. The corresponding gravitational potential is of Plummer form with the same radial scale $a$:
\begin{align}
	\Phi_d (R)= -\frac{GM_d}{\sqrt{a^2 + R^2}}\quad.
	 \label{ppa}
\end{align}

Numerical modeling of collisionless systems requires gravity softening. It can be achieved, for example, by the replacement of the Coulomb interaction law
by Plummer law:
\begin{align}
	\frac 1R \quad\to\quad \frac1{(R^2+\ve^2)^{1/2}}\quad,
\end{align}
with a cutoff parameter $\ve$. The disk gravitational potential will be also of Plummer form:
\begin{align}
	\Phi^\ve_d (R) = -\frac{GM_d}{\sqrt{b^2 + R^2}}\quad,
	 \label{pp}
\end{align}
but with the radial scale $b = a + \ve$. The form of the potential is preserved if one adds a passive spherical component of mass $M_h$ with a density
\begin{align}
\rho_0(r) = \frac{3M_h}{4\pi b^3} \left(1 + \frac{r^2}{b^2} \right)^{-5/2}\quad,
\end{align}
($r$ is a spherical radius). A full potential $\Phi^\ve$ then requires change of $M_d$ to $M = M_d + M_h$ in expression (\ref{pp}). 

Radial density distribution (\ref{sd}) can be obtained with variety of distributions in phase space. Here we use the Kalnajs DF (Kalnajs 1976). For this we
introduce new dimensionless variables for the energy $E$ and potential $ \Phi^\ve $:
\begin{align}
e \equiv E/\Phi^\ve(0) \quad,\quad w  \equiv \Phi^\ve/\Phi^\ve(0) = \frac{b}{\sqrt{b^2 + R^2}}\quad,
\end{align}
and for the angular momentum $L$ \footnote{In axisymmetric system the integral is a component of the angular momentum along the axis of symmetry. In
this article, we are only interested in this component. So, for the sake of brevity, we shall call it an angular momentum, although it is not entirely
correct.}, and the radius $R$:
\begin{align}
x \equiv -(-2E)^{1/2} L/(R_*\Phi^\ve(0)) \quad,\quad y  \equiv R \Phi^\ve(R)/(R_*\Phi^\ve(0))= \frac{R}{\sqrt{b^2 + R^2}}\quad,
\end{align}
where $R_* \equiv \lim\limits_{R\to\infty} R\Phi^\ve(R)/\Phi^\ve(0) = b$.

The surface density (\ref{sd}) can be represented as a function of two variables in the form
\begin{align}
 \sigma(w,y) = w^{m_\textrm{K}} \tau(y)\quad,
 \label{sf}
\end{align}
where $m_\textrm{K}>0$ is a free parameter of the model, while function $\tau(y)$ has a form:
\begin{align}
 \tau(y) = \frac{M_d}{2\pi a^2} (1-y^2)^{(3-m_\textrm{K})/2} (1+\delta y^2)^{-3/2}\quad,\quad \delta \equiv b^2/a^2-1\quad.
 \label{tau}
\end{align}
A DF in phase space for models in the form (\ref{sf}) can be presented as follows
\begin{align}
f(e,x) = e^{m_\textrm{K}-1} g(x) \quad,\quad M_d = (2\pi)^2\int  \frac{dE dL}{\Omega_1(E,L)} f(e,x)\quad,
 \label{df}
\end{align}
where $\Omega_1(E,L)$ is a frequency of radial oscillations. For zero cutoff parameter (\ref{tau}) coincide with (31) of Kalnajs (1976) 
and analytic expression (as a series) for the DF is possible. For $\ve>0$, the analytic expression can not be obtained and one must use
the integral representation:
\begin{multline} 
 g(x) = \{ -(2/x)^{m_\textrm{K}} / [2\pi\Phi^\ve(0)] \} \left\{ P_{m_\textrm{K}-1}(1) x\frac{\p}{\p x} [x^{m_\textrm{K}} \tau(x)] - \right.\\ 
 \left.- P'_{m_\textrm{K}-1}(1)x^{m_\textrm{K}}\tau(x) + \il_0^1P''_{m_\textrm{K}-1}(\eta)
 (\eta x)^{m_\textrm{K}} \tau(\eta x) d \eta \right\} \quad.
 \label{gx}
\end{multline}
Primes denote derivatives of Legendre functions $P_{m_\textrm{K}-1}$; $P_{m_\textrm{K}-1}(1) = 1$, $P'_{m_\textrm{K}-1} =
m_\textrm{K}(m_\textrm{K}-1)/2$. 

DF (\ref{df}) presumes all stars to rotate in the same direction, $L>0$. The surface density of the disk remains unchanged, if direction of rotation for
some stars with angular momentum $ L <L_c $ is changed. A DF with retrograde stars can be written as:
\begin{align} 
\tilde f(E,L) = \left\{\begin{array}{ll}
                 0 \quad,\quad & L<-L_c\quad,\\ [2mm]
                 \displaystyle\frac14 (2 + 3L/L_c-(L/L_c)^3) f  \quad,\quad & |L|<L_c\quad,\\ [4mm]
                 f \quad,\quad & L>L_c\quad,\\ 
                 \end{array}
          \right.
 \label{fdf}
\end{align}
where $f$ stands for $f(e, x)$ expressed through coordinates and velocities.

Thus, Kalnajs models of Kuzmin--Toomre disk have three parameters ($ m_\textrm {K}, q, L_c $): parameter $ m_\textrm {K} $ characterizes the dynamic heating
of the disk, $ q \equiv M_d / M $ is a fraction of the disk mass to the total mass in the system disk + halo, $ L_c $ regulates the amount of retrograde stars.

\subsection{The $N$-body model}

Study of disk modes within the galaxy plane and subsequent nonlinear evolution of bars requires a two-dimensional scheme in which
motion of particles is limited to galactic plane. Consequently, the position of particles in the phase space is characterized by its polar coordinates $ R,
\varphi $ and velocities $ v_R $ and $ v_\varphi $. Masses of particles are considered to be identical and equal to $ M_d / N $, where $ N $ is a number of
particles involved in a simulation.

The usual procedure for selecting the initial state involves random selection of energy $ E $ and angular momentum $ L $ for particles
according to the DF
\begin{align} 
F(E,L) = (2\pi)^2 \frac{\tilde f(E,L)}{\Omega_1(E,L)}
 \label{dF}
\end{align}
and then random selection of radius and azimuth. 
Prepared in this way, the model will have a fairly high level of initial perturbations when number of particles $ N\sim 10 ^ 5$.
This means a short linear stage of the evolution insufficient for frequency analysis. More optimal way to distribute particles in the phase space is to choose
values of $E$ at equal increments of the mass integrated over $L$, and then values of $L$ at equal increments of the mass along the cut at that $E$. Then $N_R$
equally spaced particles with the same radial and azimuthal velocities are places along the circle. 

If force calculation in the azimuthal direction is limited only to the first $ N_R-1 $ harmonics, at the initial stage these forces will be at the level
of machine roundoff error. This scheme provides a \textit{quiet start} for evolution of nonradial perturbations (Sellwood and Athanassoula 1986) and allows
for a long linear growth of unstable modes.

A $PM$-scheme with FFT is used for potential and force calculation. Spatial grid in polar coordinates ($ R $, $ \varphi $) has $ n_\varphi $ evenly spaced nodes
in azimuth and $ n_R $ nodes along the radius with concentration towards the center (Pfenniger and Friedli, 1993).

For integration of stellar trajectories we use a usual leap-frog scheme. Accuracy of the initial distribution is checked by calculation of the virial ratio of the
kinetic and potential energies, which should be close to $ -1 / 2 $ in the limit $ \ve \to 0$. Also, the system must stay in equilibrium for a long time if the
tangential force is set equal to zero. Accuracy of time integration is controlled by calculation of integrals of motion -- the energy
and the angular momentum.

\section{Appearance of a bar}

This section contains details of the evolution of Kalnajs model ($m_\textrm{K}=6$, $q=1$, $L_c = 0.25$). The model is well known in
literature and was studied by many authors. It is characterized by a flat profile of Toomre parameter  $ Q \simeq 1.5 $ (Athanassoula and
Sellwood, 1986), and moderate value of Julian--Toomre parameter (1966):
\begin{align} 
  X\equiv \frac{R\kappa^2(R)}{2\pi m G \Sigma(R)} \lesssim 3 \quad, \quad R \lesssim 5 \quad.
\end{align}
Parameters $Q$ and $X$ determine efficiency of swing amplification mechanism (Toomre 1981). 
The Plummer gravitational potential gives the slow growth of the rotation curve in the center, so the ILR doesn't interfere the passage
of the wave packets. Therefore, this model is ideal for the emergence of swing-amplified  unstable modes.

\subsection{The normal bar-mode}

\begin{figure}[b!]
\centerline{%
 \includegraphics[width=50mm, draft=false]{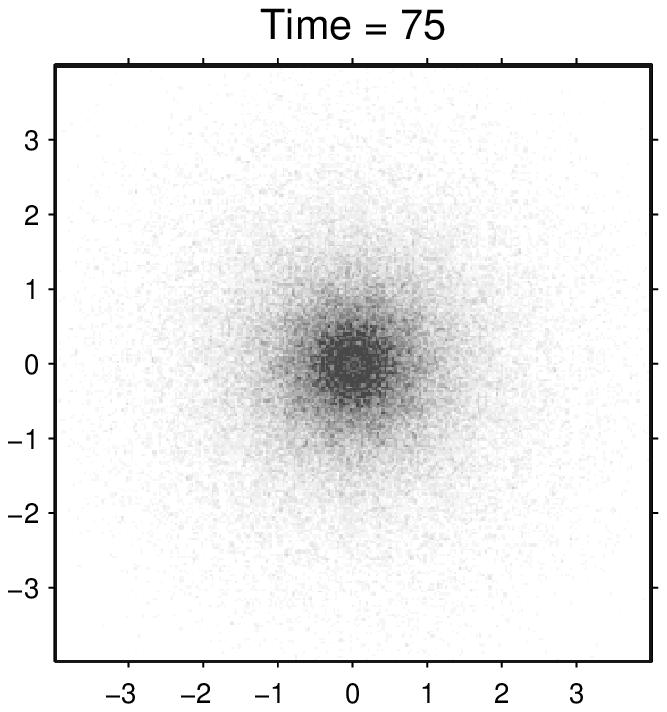}
 \includegraphics[width=50mm, draft=false]{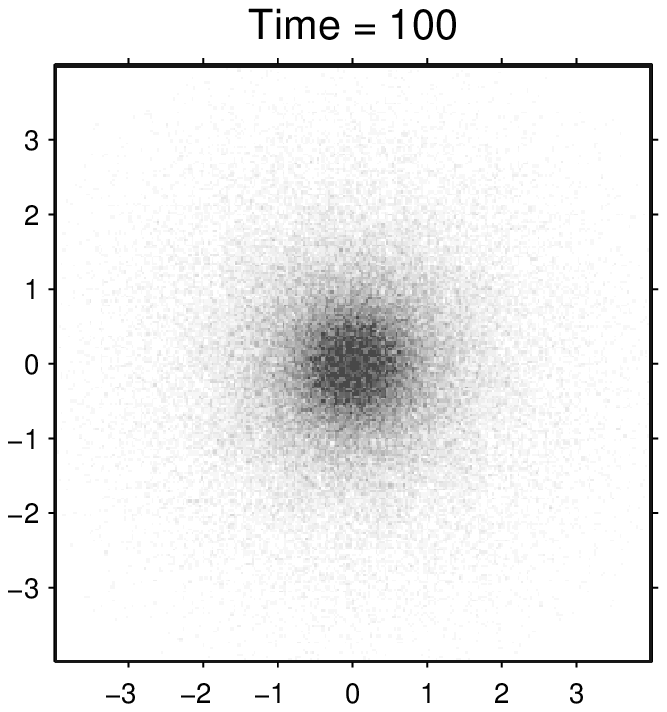}
 \includegraphics[width=50mm, draft=false]{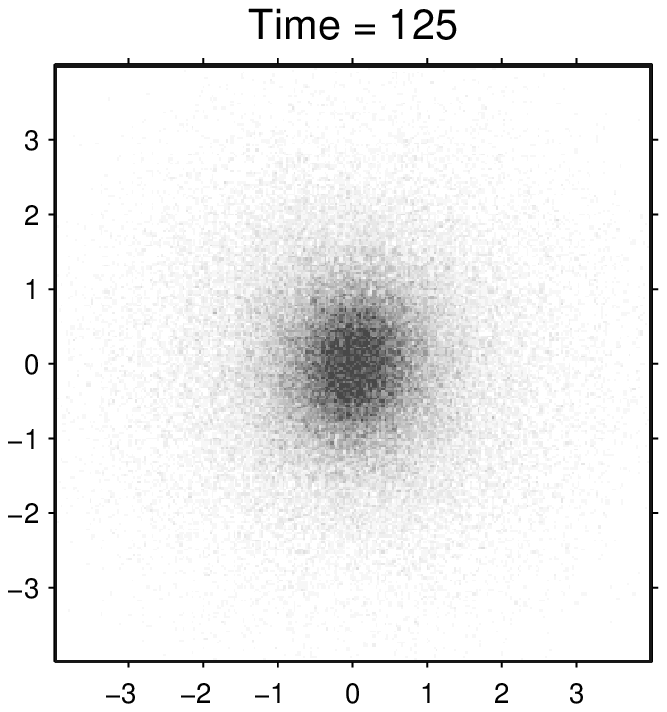} 
 }
\centerline{%
 \includegraphics[width=50mm, draft=false]{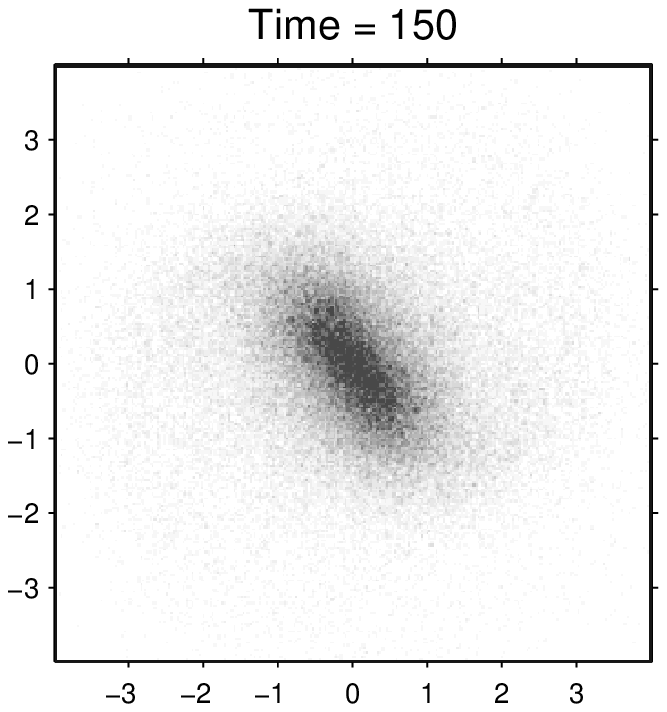}
 \includegraphics[width=50mm, draft=false]{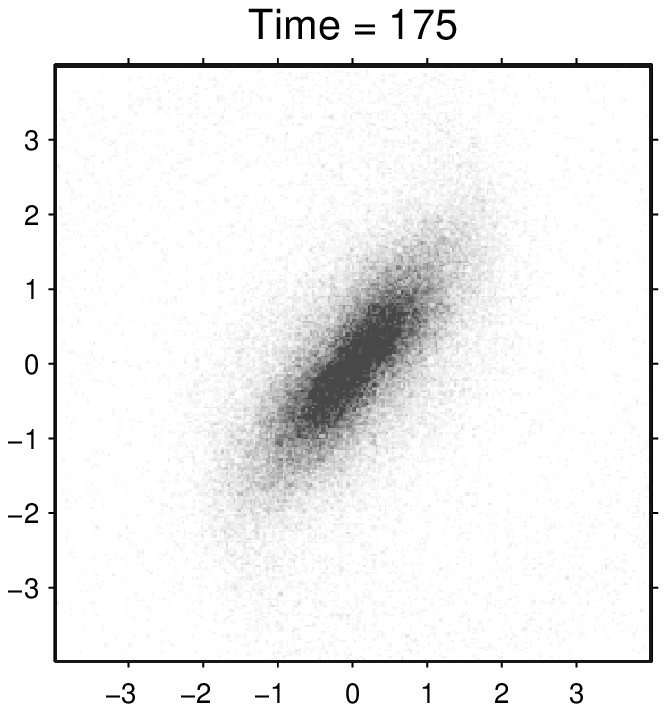}
 \includegraphics[width=50mm, draft=false]{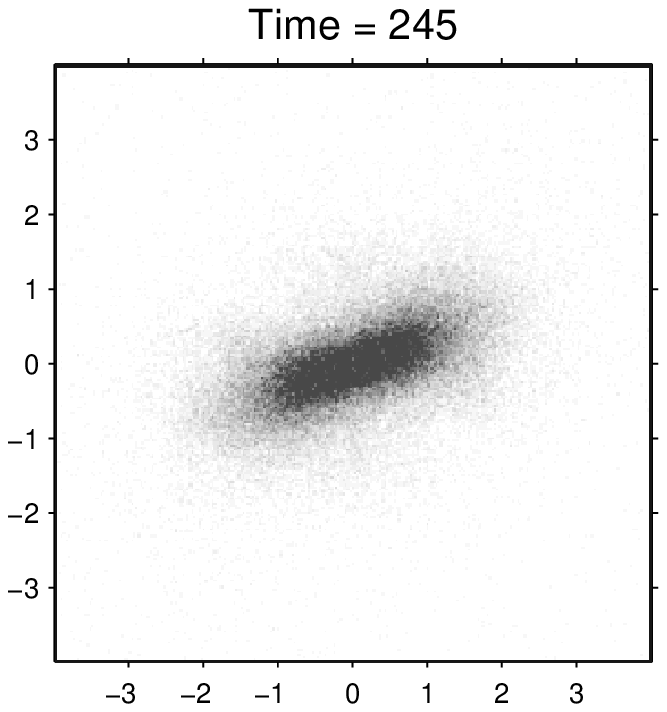} 
 }
\caption{\footnotesize Frames of the evolution of the Kalnajs (6, 1, 0.25) model with cutoff parameter $\ve = 0.2$, for moments $t=75, 100, 125, 150, 175, 245$.
Number of particles $N=10^5$. For $a=1$\,kpc and $M=10^{11} M_{\odot}$ the last frame $T=245$ corresponds to $\sim 2.45$ Gyr.}
 \label{nbody}
\end{figure}

Fig. \,\ref{nbody} shows frames of evolution of the model with cutoff parameter $\ve = 0.2$. It is assumed that $a=M=G=1$, that gives for time units
$ \simeq 10 $ $(a/1\,\textrm{kpc})^{3/2} (M/10^{11} M_{\odot})^{-1/2} $ Myr. The grid was $n_R = n_\varphi = 64$, step of time integration $h=0.024533$,
Fourier expansion of azimuthal forces was cut at $N_h=9$, number of particles on circles $N_R = 10$. Accuracy of energy and angular momentum conservation was 
$\delta_E \sim 10^{-4}$ and $\delta_L \sim 10^{-7}$. Full time of the evolution $T_{\max} = 245.33$.

The evolution of the stellar disk can be divided visibly into three stages. In the first stage $ t \lesssim 120 $
the image of the disk remains practically unchanged preserving an axisymmetric form, as in frames $ t = 75, 100$. This is followed by a relatively short second
stage
of formation of a pattern (bar-mode). The third stage $ t \gtrsim 150$ is characterized by strongly non-linear bar evolution, when bar changes slowly in length
and its rate of rotation.

A more detailed view of what is happening in the disk at these stages can be obtained by Fourier analysis and the expansion in logarithmic spirals
(Sellwood and Athanassoula 1986). Consider the expansion coefficients
\begin{align}
A_m(p,t) = \il_0^{R_{\max}} \il_0^{2\pi} \Sigma(r,\varphi,t) \exp(-i[m\varphi + p\ln r]) r
\d r \d \varphi\quad.
\end{align}
The pitch angle of log spirals is constant and equals $\mathop{\rm arctg}(m/p)$, where $p$ is spirality (trailing spirals have positive spirality,
bars correspond to $p=0$). In what follows, we are interested only in the two-arm perturbations, i.e. $ m = 2 $. Having the $ N $-body model surface density  in the
form
\begin{align}
\Sigma(r,\varphi,t) = \frac{1}{Nr} \sum\limits_i \delta[r-r_i(t)] \, \delta[\varphi-\varphi_i(t)]\quad,
\end{align}
where $r_i(t)$ и $\varphi_i(t)$ are coordinates of particles, one obtains
\begin{align}
A_m(p,t) = \frac{1}{N} \sum\limits_i \exp(-i[m\varphi_i + p\ln r_i])\quad.
\end{align}

\begin{figure}[tb!]
\centerline{%
 \includegraphics[width=80mm, draft=false]{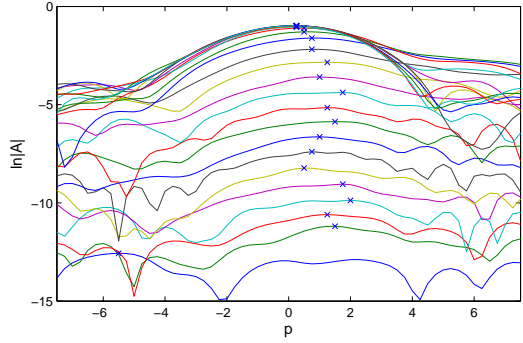}
 }
\caption{\footnotesize The dependences of $\ln|A_2(p,t)|$ v.s. spirality $p$ shown for different moments of evolution $t=0, 10, 20, ..., 240$. Crosses show maxima
for each curve.}
 \label{Apt}
\end{figure}

Fig.\,\ref{Apt} shows the curves $|A_2(p,t)|$ characterized by rather broad plateau, indicating the presence of components with
different $ p $ in the perturbation. 
At the beginning the plateau is localized in the interval $-1<p<3$ with average spirality $p=1$.
Then, in the course of the pattern development, it moves left in the interval $ -2 <p <2 $ with the average value $ p = 0 $. Crosses, indicating the positions of
the maxima of the curves also show a tendency of the pitch angle increase with growth of the disturbance amplitude. It is important,
however, that function $ | A_2 (p, t) | $ has no pronounced maxima at large and opposite values of $p$ corresponding to leading and trailing tightly-wound spirals,
as can be expected in the Toomre theory of bar instability.

\begin{figure}[b!]
\centerline{%
 \put(0, 120){a)}\hspace{2mm} \includegraphics[width=80mm, draft=false]{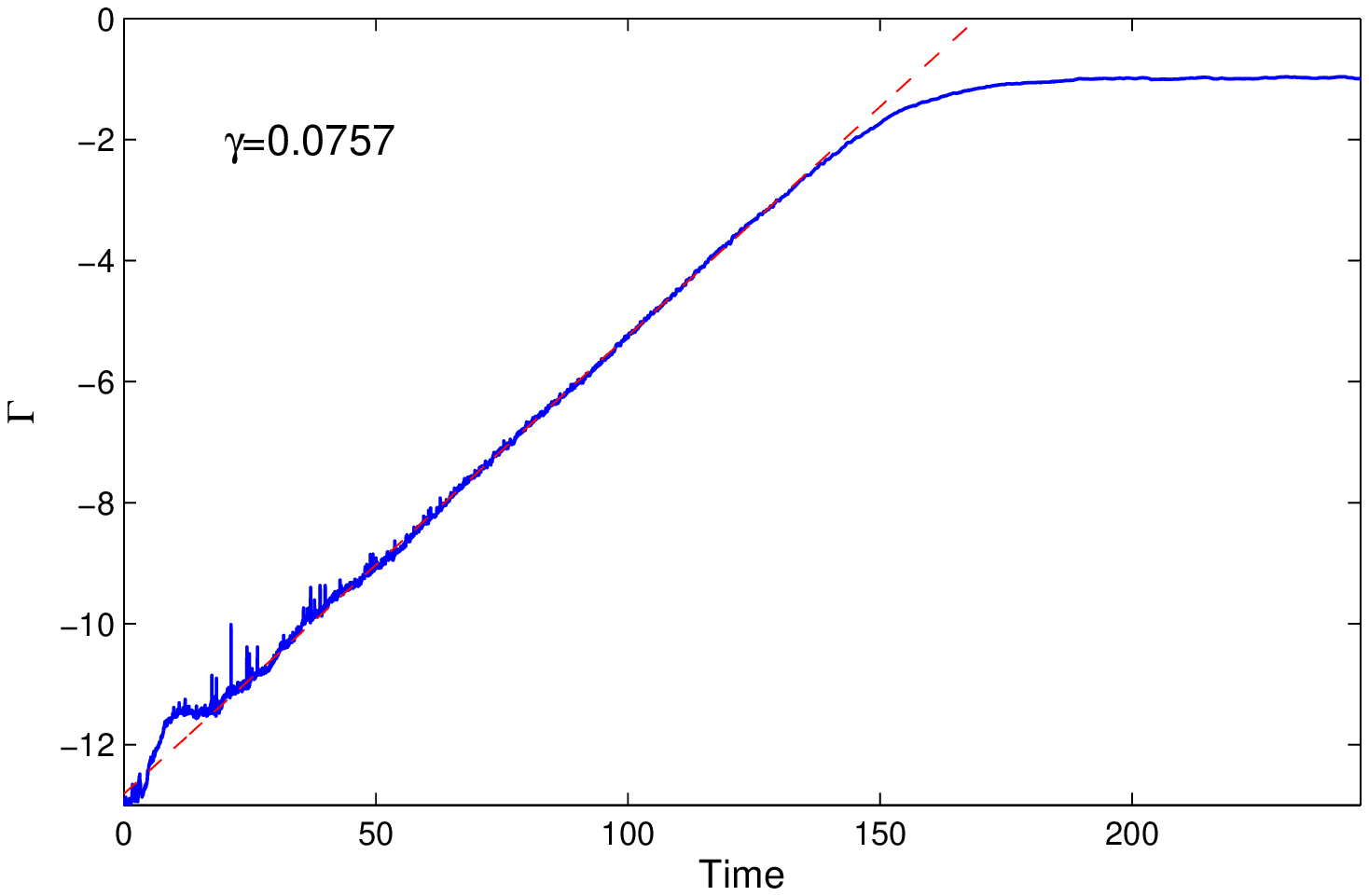}
 \put(0, 120){b)}\hspace{2mm} \includegraphics[width=80mm, draft=false]{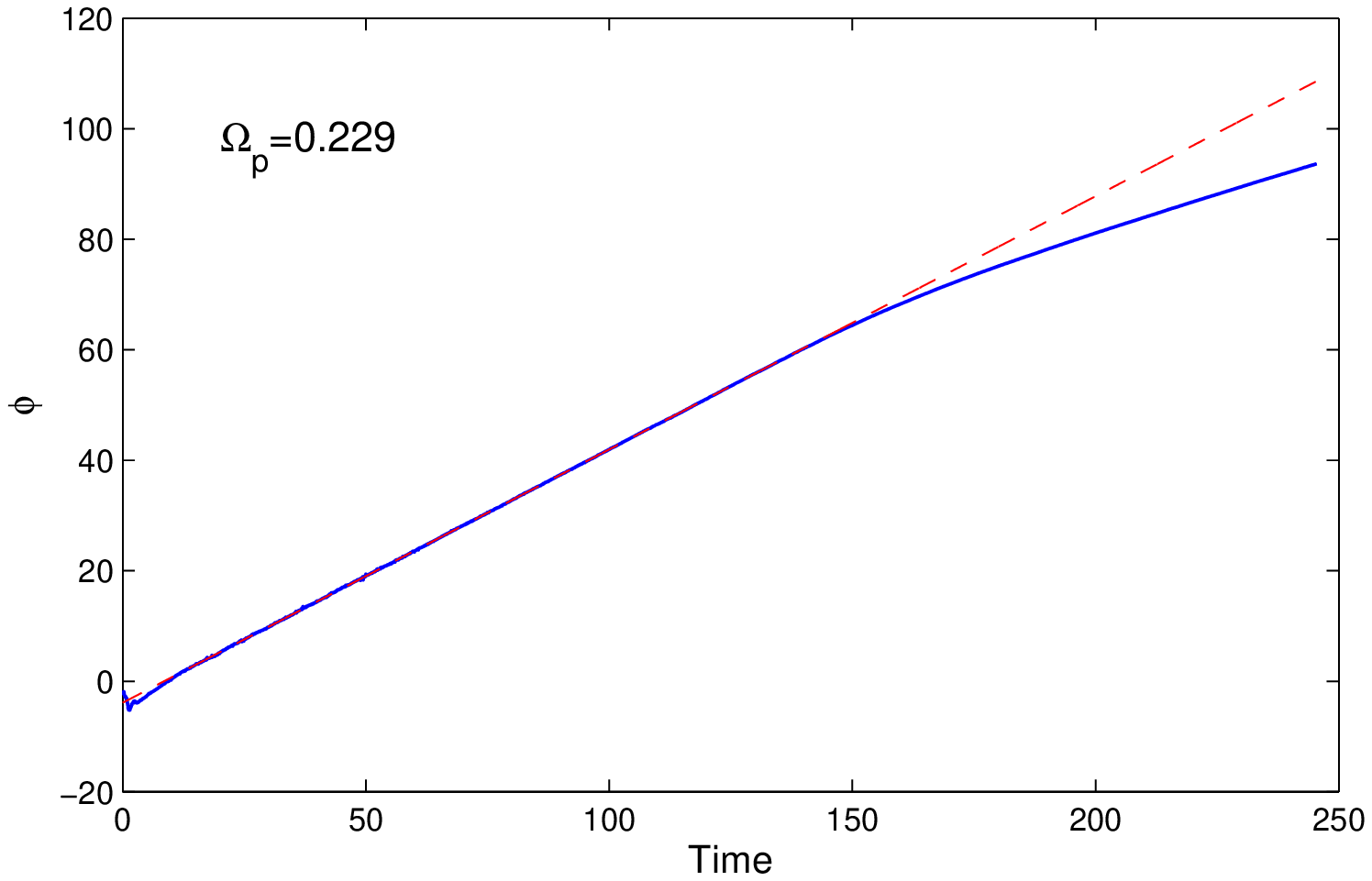} 
 }
\caption{\footnotesize Time dependences of $\Gamma(t) = {\mlp} \ln|A_2(p,t)|$ (a) and complex phase $\phi$, $A_2(p,t) = |A_2| e^{i\phi}$ (b) found at fixed $t$ for
values of $p$ for which $\ln|A_2(p,t)|$ peaks. Dashed lines show linear behavior for both dependences obtained by least square method for $10<t<130$.  }
 \label{OmG}
\end{figure}

Fig.\,\ref{OmG} shows time dependence for $\Gamma(t) = {\mlp} \ln|A_2(p,t)|$ and complex phase $\phi(t)$, $A_2(p,t) = |A_2| e^{i\phi}$ found at fixed $t$ for values
of $p$ for which $\ln|A_2(p,t)|$ peaks. The smoothness of the curves and the presence of broad plateau in Fig.\,\ref{Apt} indicates that the result should not
depend on a particular form of spirals used for decomposition.

The main thing to note in Fig.\,\ref{OmG}a that curve $ \Gamma (t) $ follows precisely the linear law at $ t<140 $, up to amplitudes $ \ln | A_2 |
\simeq -2.3 $. The linearity means that the amplitude of the perturbation grows exponentially with time, i.e. $ \propto \exp (\gamma t) $, where $ \gamma $ is the
growth rate. The slope of the linear region of $ \Gamma (t) $ determines the growth rate of the instability. After reaching a level of about the
unperturbed surface density at $ t \sim 170$, the bar amplitude saturates at the level $ \ln | A_2 | \simeq -1 $.

The phase $\phi(t)$ demonstrates even better linear behavior at $t<150$, see Fig.\,\ref{OmG}b, which means a constant rate of rotation of the global pattern
\begin{align}
\Omega^\ve_p = \frac1m \frac{d \phi}{d t}\quad .
\label{omp}
\end{align}
After the linear stage, the rate of rotation slows down. 

A fairy long period of linear evolution $ t < 140$ with constant rotation velocity and growth rate suggests the existence of an unstable
normal mode. Note that no sign of the pattern can be visibly detected for most of this period (up to about $ t <117 $, $ \ln | A_2 | \lesssim -4 $), and a quite
well shaped bar appears only at the end of the linear phase at $ t \approx 150 $.

\begin{table}[b!]
\centering
\footnotesize
\newcolumntype{C}{>{$}c<{$}}
\begin{tabular}{*{7}{C}ccc}
\toprule	
\ve       & 0.20 & 0.15 & 0.10 & 0.07 & 0.05 & & 0 & & JH \\
\midrule
\Omega_p  & 0.229& 0.267 & 0.314 & 0.344 & 0.366 & & 0.428 & & 0.445 \\
\midrule
\gamma    & 0.076& 0.110 & 0.161 & 0.196 & 0.229  & & 0.312 & & 0.308 \\
\bottomrule	
\end{tabular}
\caption{Pattern speeds $\Omega_p$ and the growth rates $\gamma$ of the mode found from slopes of $\Gamma(t)$ and $\phi(t)$ at values of cutoff parameter $\ve =
0.2$, 0.15, 0.1, 0.07, 0.05 and its extrapolation to $\ve=0$. The last column contains figures of independent linear perturbation analysis by Jalali and Hunter
(2005) for the model without softening.}
\label{tab1}
\end{table}

\begin{figure}[tb!]
\centerline{%
 \put(0, 120){a)}\hspace{2mm} \includegraphics[width=80mm, draft=false]{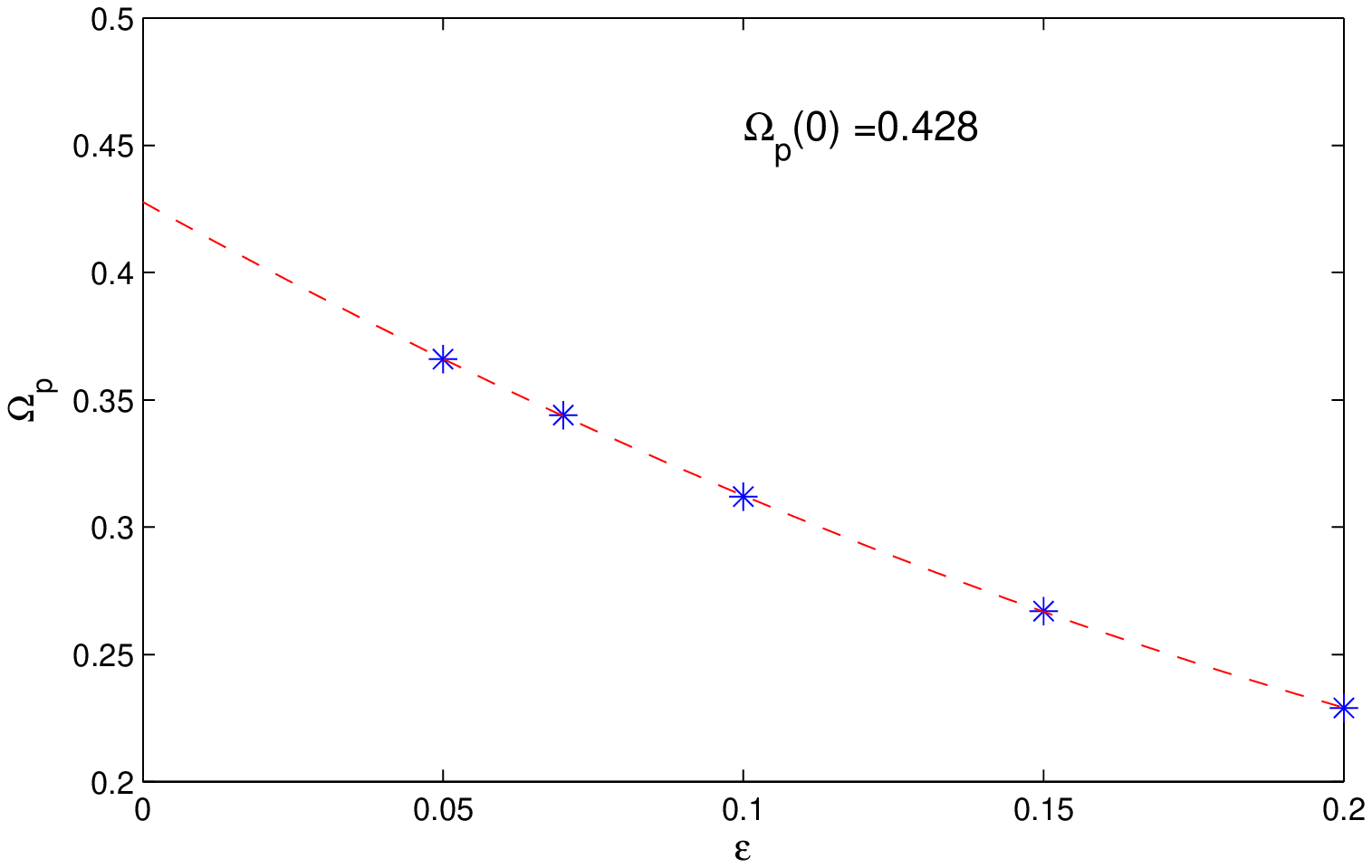}
 \put(0, 120){b)}\hspace{2mm} \includegraphics[width=80mm, draft=false]{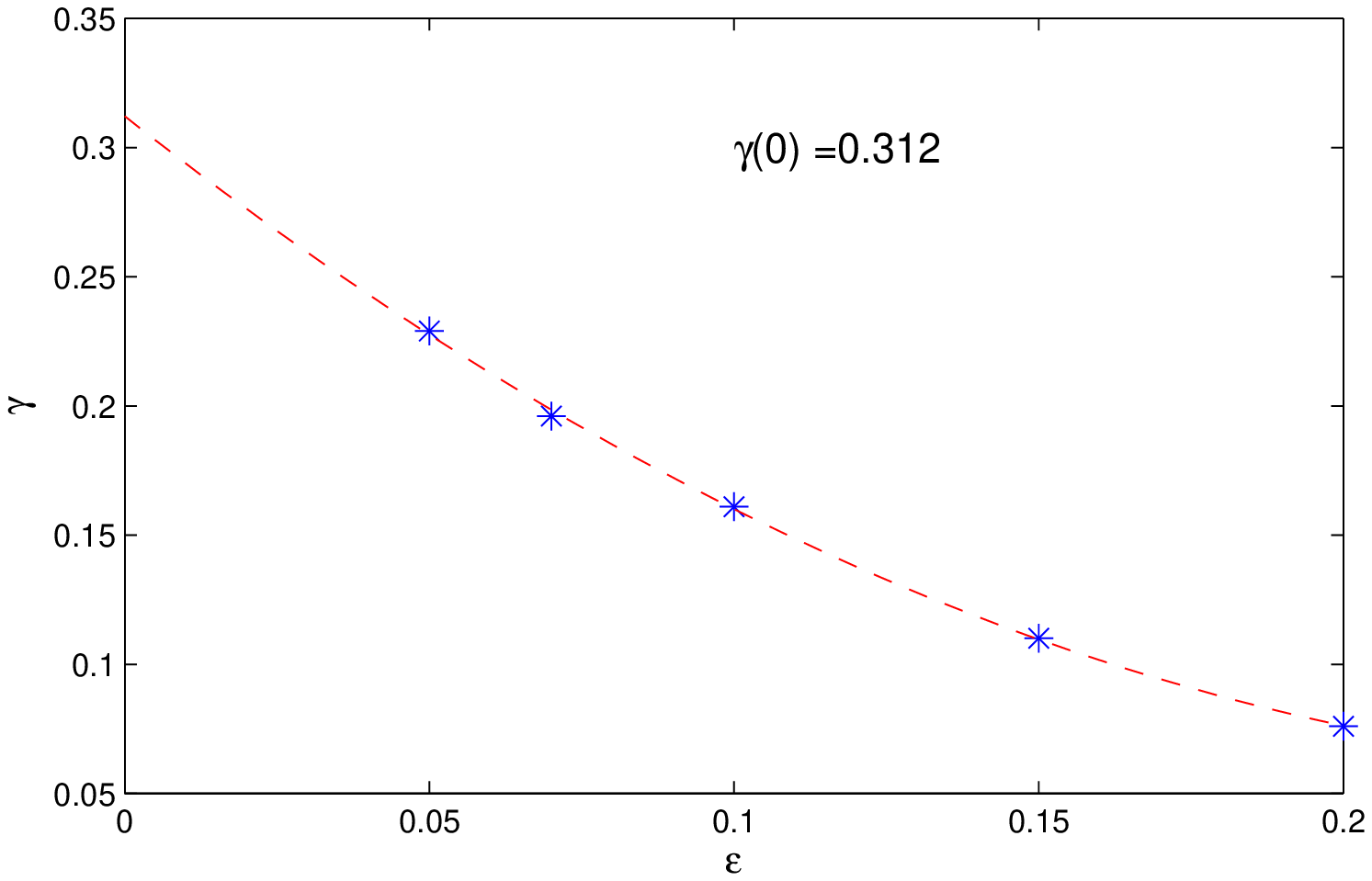}
 }
\caption{\footnotesize Pattern speeds $\Omega_p$ and the growth rates of the mode $\gamma$ at values of cutoff
parameter $\ve = 0.2$, 0.15, 0.1, 0.07, 0.05 and its extrapolation to $\ve=0$.}
 \label{lim}
\end{figure}

The pattern speeds and growth rates calculated for different values of the cutoff parameter $ \ve $ are shown in
Fig.\,\ref{lim} and in the table. The least square quadratic extrapolation to $ \ve = 0 $ gives for the pattern speed without softening $ \Omega_p^0 \simeq
0.428 $, and for the growth rate $ \gamma^0 \simeq 0.312 $.

Jalali and Hunter (2005) independently investigated the model without softening by linear perturbation analysis. They have found two unstable normal modes. For the
most unstable mode, the pattern speed $ \Omega_p \simeq 0.445 $, and the growth rate $ \gamma \simeq 0.308 $, for the second unstable mode -- $ \Omega_p \simeq
0.294 $, and the growth rate of $ \gamma \simeq 0.109 $. Our values are very close to the values of the first mode found by Jalali and Hunter. This once again
confirms the conclusion that we are dealing with the most unstable normal mode.

A shape of the mode remains unchanged in the linear stage of the evolution. It looks like an open
spiral with one maximum (see Fig. 3a in Jalali, Hunter 2005). In the next section we will see how this unstable bar mode is transformed
into a bar in the nonlinear stage.

\subsection{Nonlinear evolution of the bar mode}

The numerical experiment demonstrates the presence of one unstable mode. The developed bar on the late stage of the evolution is a nonlinear structure with an
amplitude comparable with the axisymmetric background. Examination of the power spectra of disk oscillations at different time intervals allows one to
associate a small amplitude unstable bar-mode and the non-linear bar. 

Let's introduce two kinds of power spectra for different spirality $p$ and radii in the interval $(t, t + s)$. First of all, for determination of pattern
speed $\Omega_p$ (or real part of the frequency $\mathrm{Re}\,\omega = m\Omega_p$) it is necessary to exclude the growth of disturbances. One way of achieving it is
to exclude the exponential growth factor $\exp(\gamma t')$ from the amplitude $A_2(p,t')$. Then the expression for the spectra will be limited to the
linear stage, while the growth rate is constant and equal to $\gamma \equiv \mathrm{Im}\,\omega$. 
A more general way is to normalize the amplitude to the maximum of its modulus $ \Gamma (t ') $, thus the expression for the spectra can be used
both in linear and nonlinear regimes, even when there is no increase in amplitude. Therefore we define the power spectrum, corresponding to different values of
spirality, by the expression:
\begin{align}
{\cal P}^s_t(p,\Omega_p) = \frac{1}{2\pi}\left| \il_t^{t + s} \frac{A_2(p,t')}{\Gamma(t')} e^{im\Omega_p t'} H_t^s(t') dt'\right|^2 \quad,
\label{power}
\end{align}
where $H_t^s(t')$ is the Hann filter:
\begin{align}
H_t^s(t') = \frac12 \left[1 - \cos\( 2\pi \frac{t'-t}{s}\) \right] \quad.
\label{hann}
\end{align}
Similarly, for the radial spectrum:
\begin{align}
{\cal R}^s_t(R,\Omega_p) = \frac{1}{2\pi}\left| \il_t^{t + s} \frac{\tilde A_2(R,t')}{\Gamma(t')} e^{im\Omega_p t'} H_t^s(t') dt'\right|^2 \quad,
\label{rom0}
\end{align}
where
\begin{align}
\tilde A_2(R,t) = \il_{R-\Delta R}^{R+\Delta R} \il_0^{2\pi} \Sigma(r,\varphi,t) \exp(-im\varphi) r \d r \d \varphi = \frac{1}{N} {\sum\limits_i}'
\exp(-im\varphi_i)\quad.
\end{align}
Prime in the last expression means summation over particles in the ring $(R-\Delta R, R+\Delta R)$.

\begin{figure}[tb!]
\centerline{%
 \put(0, 120){a)}\hspace{2mm} \includegraphics[width=80mm, draft=false]{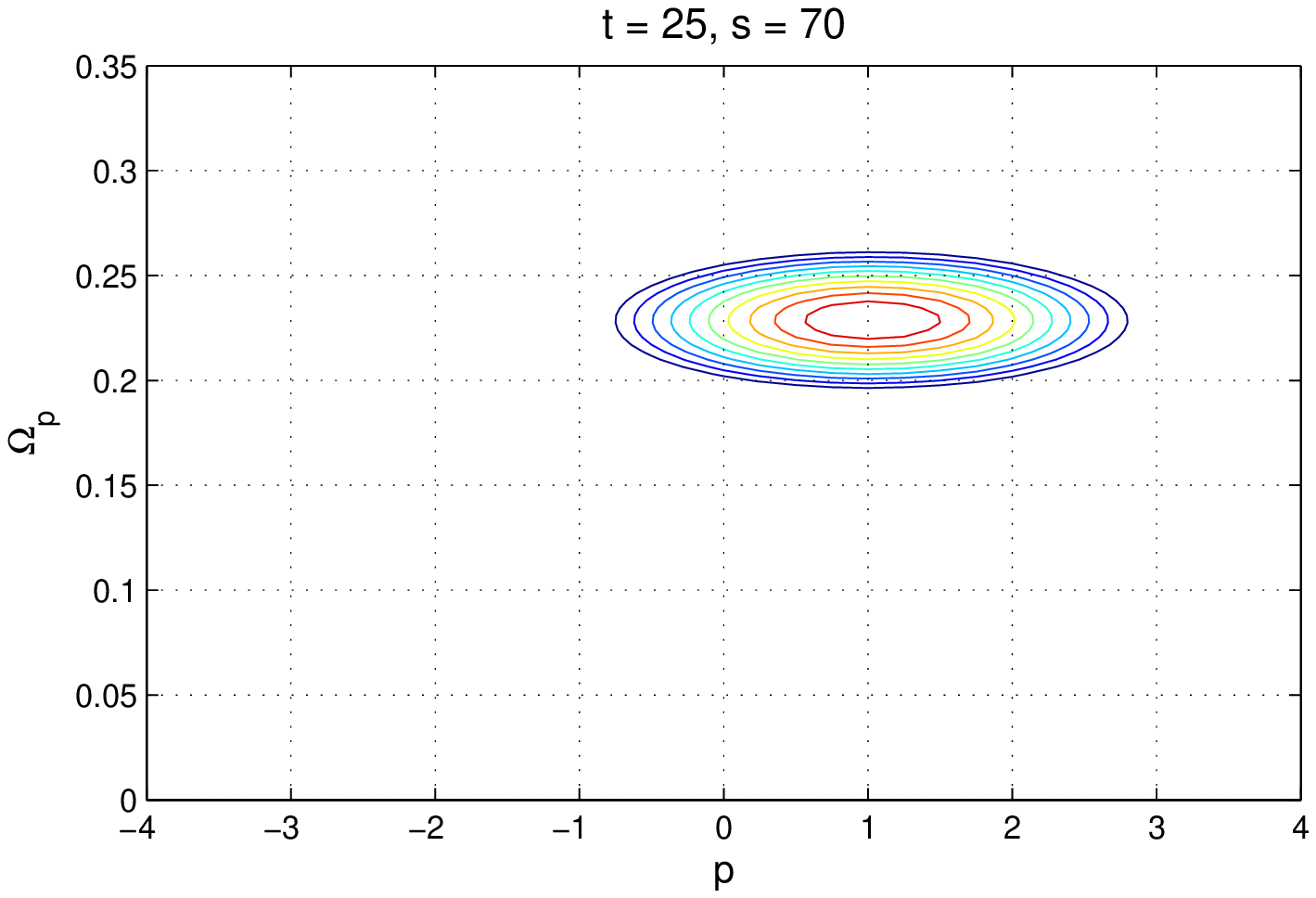}
 \put(0, 120){b)}\hspace{2mm} \includegraphics[width=80mm, draft=false]{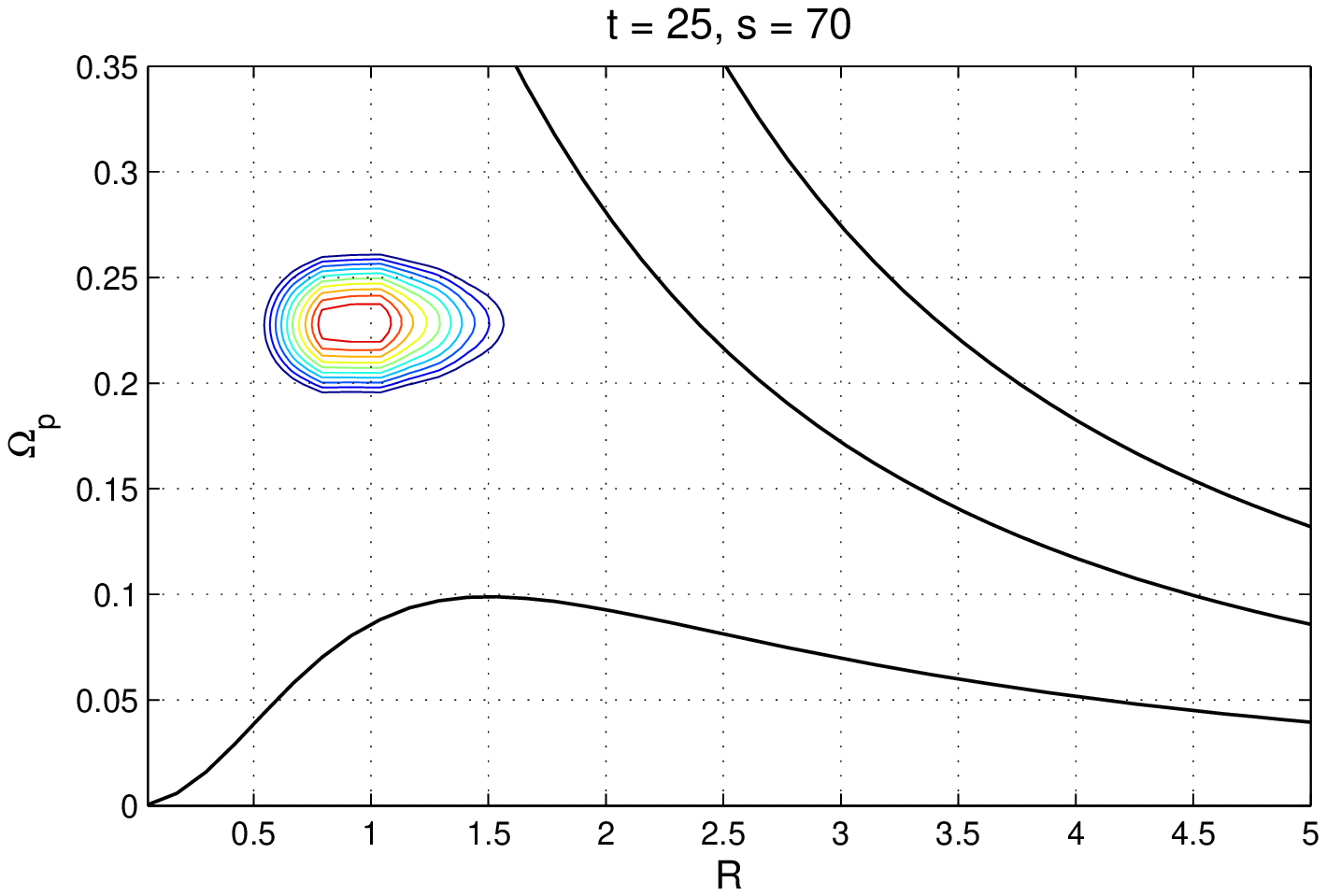}
 }
\centerline{%
 \put(0, 120){c)}\hspace{2mm} \includegraphics[width=80mm, draft=false]{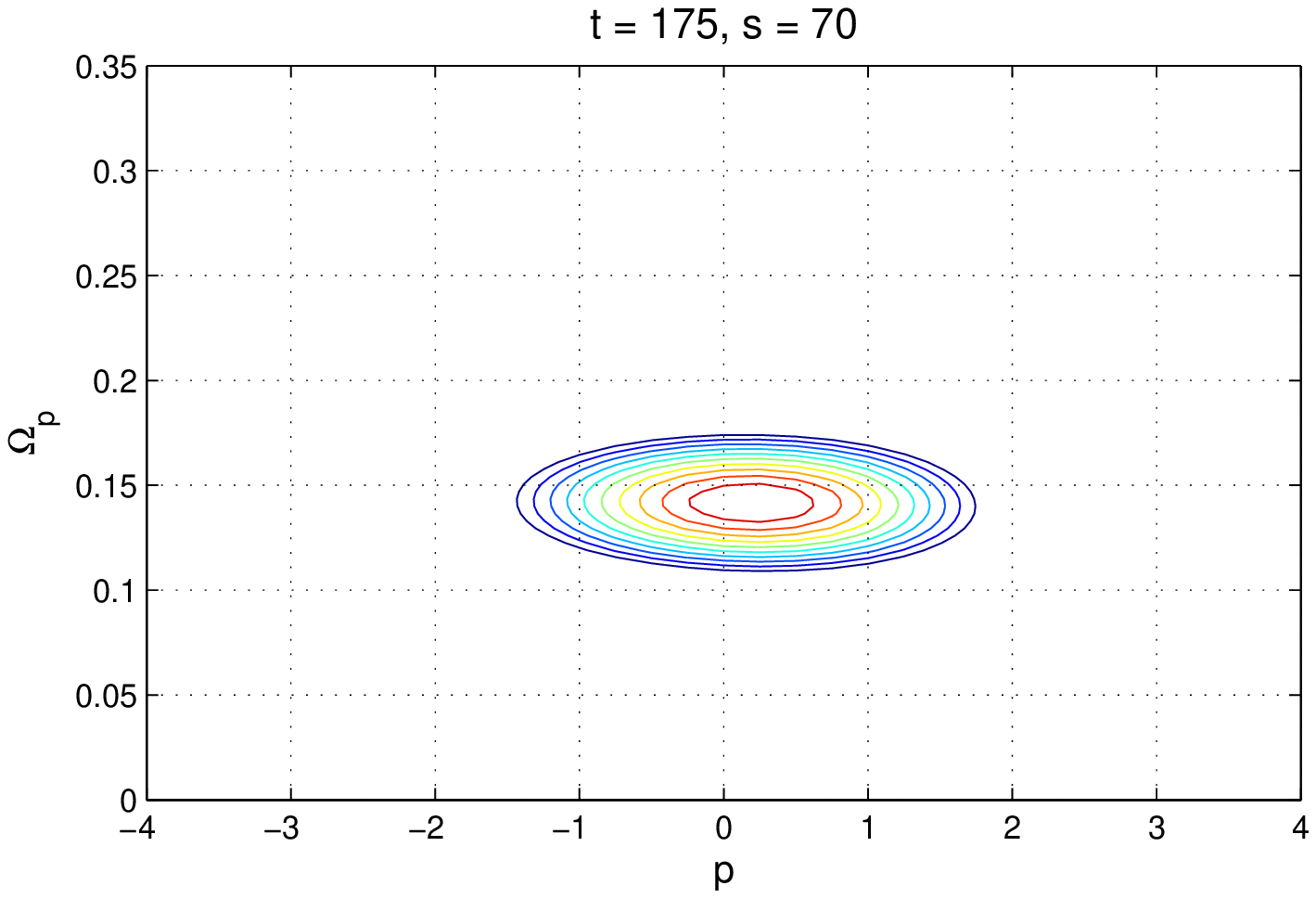}
 \put(0, 120){d)}\hspace{2mm} \includegraphics[width=80mm, draft=false]{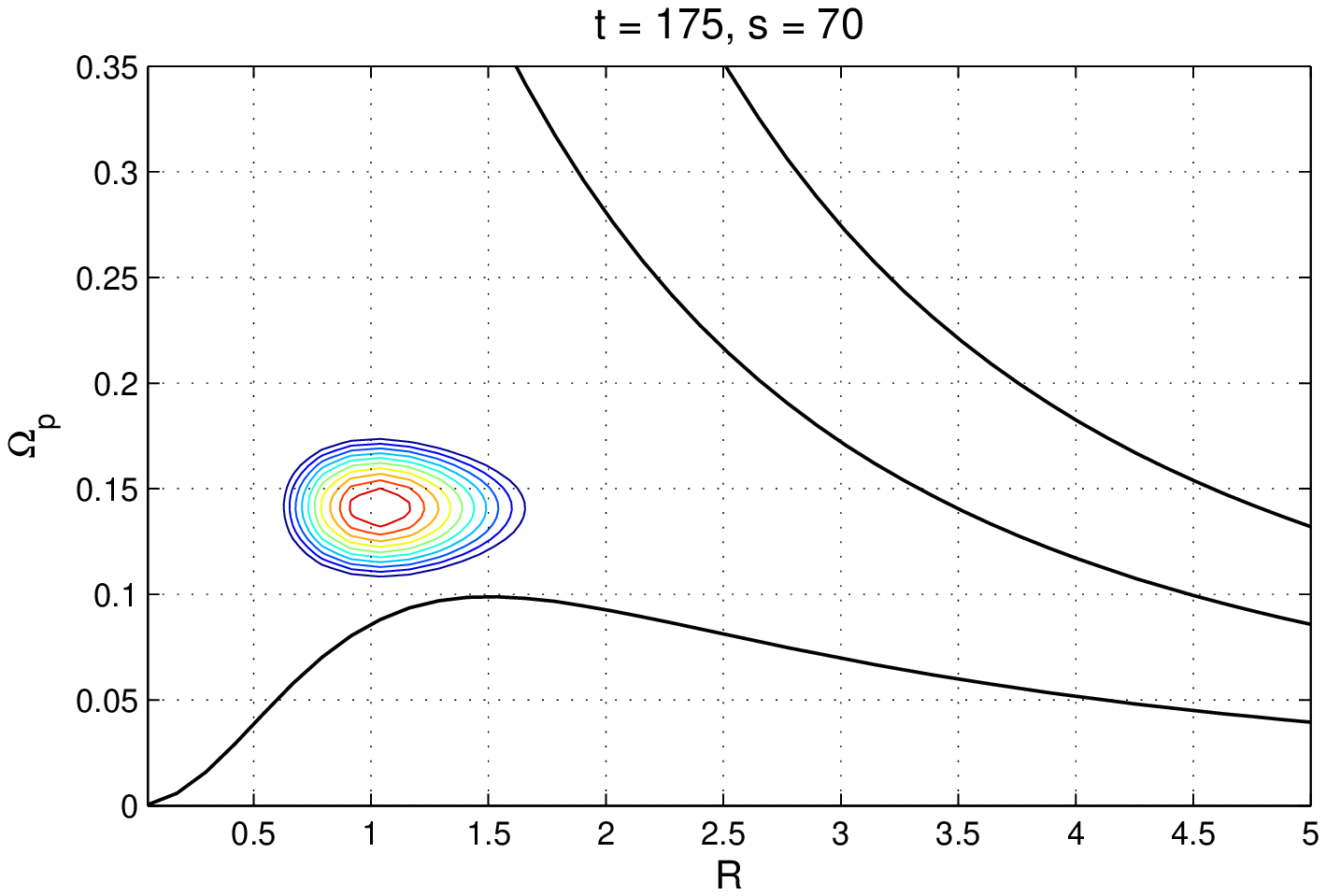}
 }
\caption{\footnotesize Isolines of the power spectra (\ref{power}) (figures a,c) and (\ref{rom0}) (figures b,d), for the linear phase of the evolution of the
bar-mode $25<t<95$, and for the nonlinear evolution of the bar $175< t < 245$. Filter width in all cases $s=70$. Outer contours on each figure correspond to half
of the spectra maxima, next contours increment 5\% of the maxima. Solid lines in figures b), d) show frequency curves $\Omega-\kappa/2$, $\Omega$,
$\Omega+\kappa/2$, similar to Fig.\,\ref{fig_res}. }
 \label{spectr}
\end{figure}

Fig.\,\ref{spectr} shows power spectra for the model with cutoff parameter $\ve=0.2$ for linear ($25<t<95$) and nonlinear ($175<t<245$) phases of bar evolution.

The peaks of the power spectra show the position of the coherent structures, in which a large number of particles are involved. Fig. a) and b)
are the power spectra corresponding to different values of $ \Omega_p $ and spirality $ p $, calculated from (\ref{power}). Similar to frames of the evolution
displayed in Fig.\,\ref{nbody}, one can trace time evolution of the power spectra. During the linear phase, contours of power spectra keep their
position near the point $\Omega_p \simeq0.23$ and spirality $p=1$, corresponding to an open spiral with an average pitch angle $\alpha = 60^\circ$. 
During the transition period and the nonlinear stage, one can observe the pattern's slowdown and its transformation into a bar (with $p\approx 0$).

Figures b) and d) show maxima of the radial power spectrum (\ref{rom0}) revealing radial localization of the mode. The frequency curves taken from
Fig.\,\ref{fig_res}, but with softening, allow one to compare the position of the spectrum maxima with the position of the maximum of the precession curve
$\Omega-\kappa/2$. It can be seen that the disturbance is concentrated over the maximum, according to the theory of bar formation by Polyachenko (2004). After the
slowdown, the bar speed becomes  slightly higher than the maximum of the precession curve.

\begin{figure}[bt!]
\centerline{%
 \put(0, 120){a)}\hspace{2mm} \includegraphics[width=80mm, draft=false]{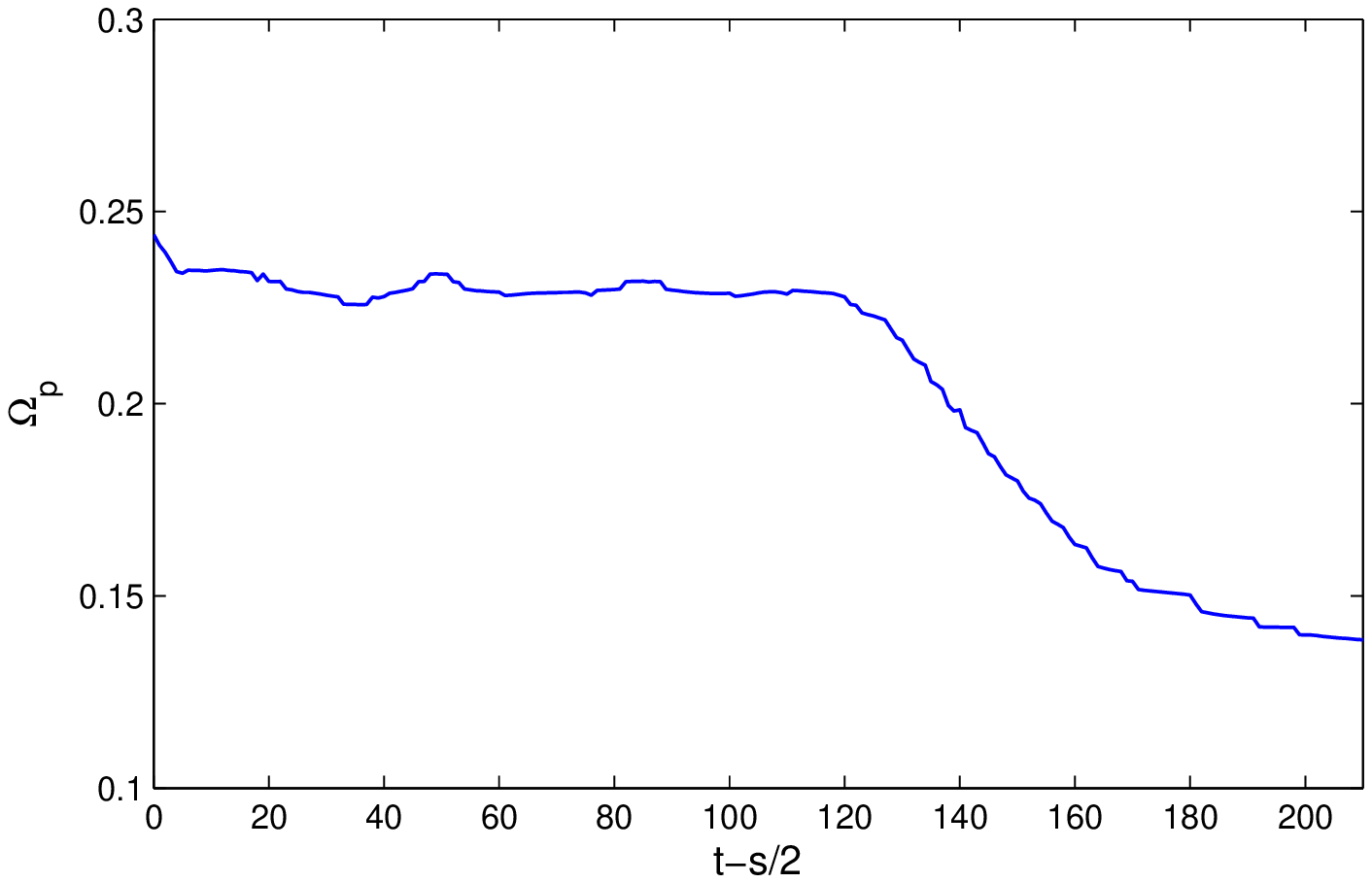}
 \put(0, 120){b)}\hspace{2mm} \includegraphics[width=80mm, draft=false]{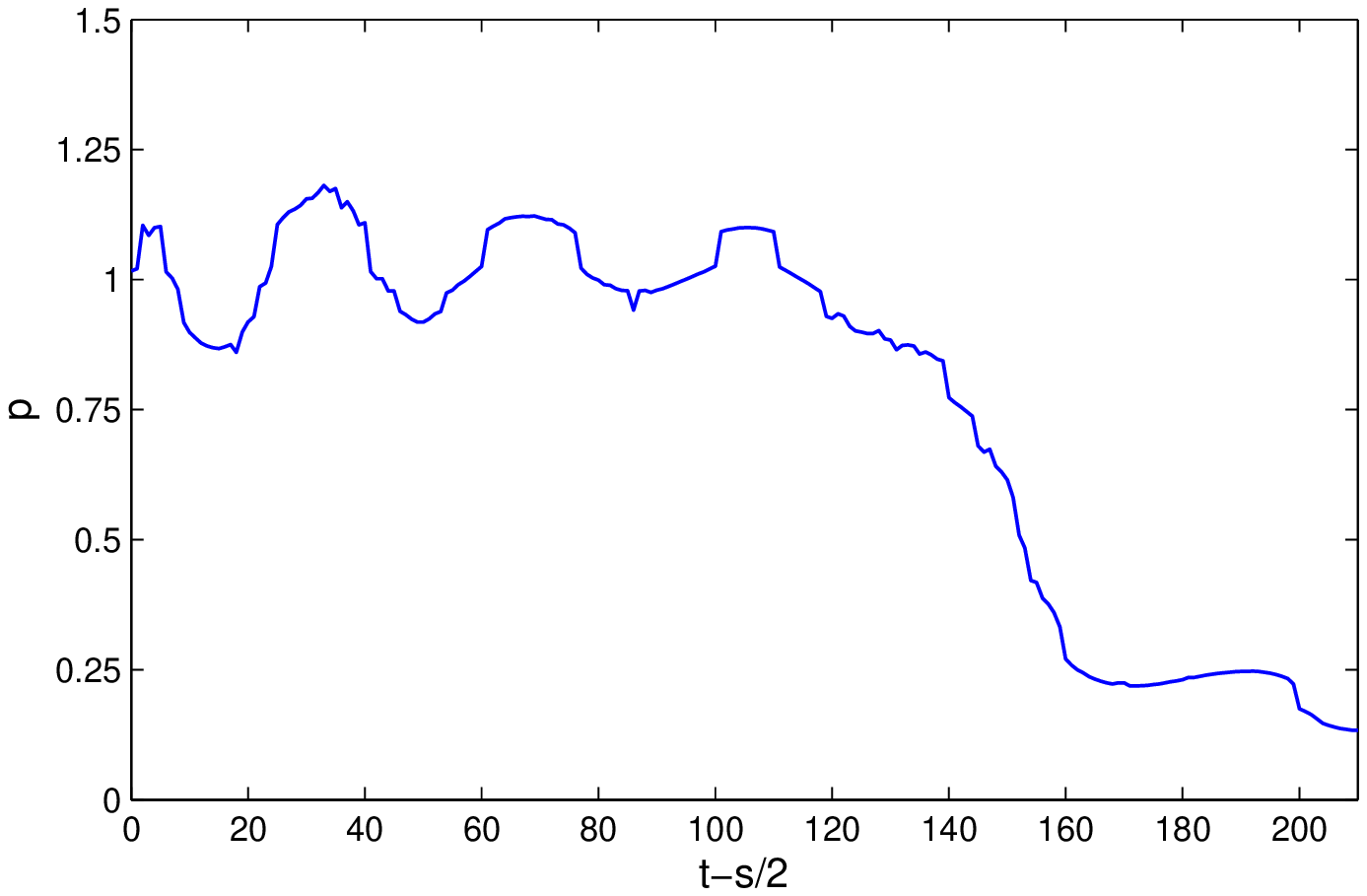}
 }
\caption{\footnotesize Time dependence of the pattern speed $\Omega_p(t)$ (a) and spirality $p(t)$ (b), found by location of power spectra maxima (\ref{power}) and
(\ref{rom0}), for the bar-mode (cutoff parameter $\ve=0.2$). The filter width is keep constant, $s=30$.  
}
 \label{spt}
\end{figure}

The described temporal behavior of the pattern speed $\Omega_p$ and spirality $p$ is clearly seen in Fig.\,\ref{spt}, that shows location of maxima of  
${\cal P}^s_t(p,\Omega_p)$ versus time. The plot in Fig.\,\ref{spt}а is consistent with our findings based on analysis of Fig.\,\ref{OmG}b,
about the behavior of the pattern speed, taking into account a time lag $s/2$ of the filter (\ref{hann}).

\section{Discussion}

The quiet start technique by Sellwood and Athanassoula (1986) allows one to construct stellar disk models with low level of initial perturbations and to
study their linear and nonlinear evolution using different Fourier spectra. 

Our investigation of Kuzmin--Toomre disk model with flat Toomre parameter profile $ Q \approx 1.5 $ reveals the presence of a disk unstable normal mode. This
mode dominates in the disk from the beginning of the simulations. A comparison with the independent results of the linear perturbation analysis by Jalali and Hunter
(2005) shows that it is the most unstable normal mode of the stellar disk, the surface density profile of which has a unique maximum.

Upon reaching the level comparable with the axisymmetric background, the mode is governed by non-linear evolution: the amplitude stops growing and the pattern speed
slows down. Using power spectra we see that the pattern preserves the open spiral shape during the linear evolution, then it transforms
into a more open spiral and eventually a bar.

Frequency analysis of perturbations hasn't shown any other unstable modes. Clearly, this is the result of our crude analysis that removes all the modes except
the most unstable one.

The model meets necessary conditions for effective swing amplification of transient waves. Thus, it would be natural to expect the presence of leading and
trailing spirals that form the bar. However, we could not find it neither visually nor using a more detailed frequency analysis. According to Binney and Tremaine
(2008), in case of considerable growth rates the swing mechanism should give spiral patterns, while bar-modes are characterized by low growth rates and lumpy
structures. We have seen that it contradicts to results of our numerical experiments, where the bar is formed from the most unstable mode with one maximum.

\bigskip

Note that in the literature there are different interpretations of bar instability even in the framework of the swing amplification. For example, in the recent
review by Sellwood (2010) one can find: <<Toomre (1981) provided the most important step forward by elucidating the mechanism of the bar instability (see also BT08
sect. 6.3). Linear bar-forming modes are standing waves in a cavity, akin to the familiar modes of organ pipes and guitar strings. Reflections in galaxies take
place
at the center and at the corotation radius, except that outgoing leading spiral waves incident on the corotation circle are super-reflected into amplified ingoing
trailing waves (i.e. swing amplification), while also exciting an \textit{outgoing transmitted trailing wave}. The feedback loop is closed by the ingoing trailing
wave reflecting off the disk center into a leading wave, which propagates outwards because the group velocity of leading waves has the opposite sign to that of the
corresponding trailing wave. The amplitude of the continuous wave-train at any point in the loop rises exponentially, because the circuit includes positive
feedback.>>

First, the reflection takes place from a $Q$-barrier located near the corotation. A wave cannot propagate freely under the barrier.
Second, the outgoing transmitted trailing wave behind the $Q$-barrier hasn't been mentioned in Toomre (1981) at all. For the first time it appeared in waser-I
mechanism by Mark (1974) for spiral disturbances. Meanwhile Sellwood (2010) repeats the so-called waser-II mechanism for bar-modes proposed by Bertin (1983), which
is the example of a modal approach, where bars are treated as unstable normal modes. 

Waves outside corotation ($\Omega<\Omega_p$) have positive angular momentum, while waves inside corotation ($\Omega > \Omega_p$) have negative angular momentum.
Penetration through the $Q$-barrier provides a mechanism for the instability since generation of positive angular momentum wave strengthens a negative angular
momentum wave and vise versa. The presence of a third wave is important, since it explains the source of instability by means of conservation laws. If
swing amplification mechanism does not require the third wave, it is quite different from what is usually meant. 

It seems, however, that Toomre's approach to bar formation in galaxies involving transient spirals and due to swing amplification is not a necessity. 
There are many reasons, especially the relatively recent observations in the near-infrared, to believe that instead barred spiral structures should be naturally
interpreted in terms of global modes. The amplification mechanism can be the same independently whether the structures are transient or not. The transient-swing
amplification picture is a mathematical description that has a direct counterpart in the dynamics of steady wavetrains at corotation.

Comparison of our theory of bar formation with waser-II mechanism requires further investigation, since both mechanisms are modal. For example, one can
investigate the effect of feedback break on the stability properties of stellar disks and what is the relative role of the corotation region, the outer Lindblad
resonance, and other external resonances in bar mode instability .

\section*{Acknowledgements}

The work was supported in part by Russian Science Support Foundation, RFBR grants No. 11-02-01248, No.~12-02-33118, and also by Programs
of Presidium of Russian Academy of Sciences No. 17  ``Active Processes in Galactic and extragalactic objects''.

\end{document}